\documentclass[12pt]{article}

\usepackage{axodraw}
\usepackage{amsmath}
\usepackage{amsfonts}

\newtheorem{lemma}{Lemma}

\begin{document}
\baselineskip=20pt

\pagenumbering{arabic}
  
\vspace{1.0cm}
\begin{flushright}
LU-ITP 2004/049
\end{flushright}

\begin{center}
{\Large\sf Retarded Functions in Noncommutative Theories}
\\[10pt]
\vspace{.5 cm}

{Tobias Reichenbach\footnote{\emph{Email adress:} tobias.reichenbach@itp.uni-leipzig.de}}
\vspace{1.0ex}

{\small Institut f\"ur Theoretische Physik, Universit\"at Leipzig,
\\
Augustusplatz 10/11, D-04109 Leipzig, Germany\\}

\vspace{2.0ex}

\end{center}

\begin{abstract}
The perturbative approach to quantum field theory using retarded functions is extended to noncommutative theories. Unitarity as well as quantized equations of motion are studied and seen to cause problems in the case of space-time noncommutativity. A modified theory is suggested that is unitary and preserves the classical equations of motion on the quantum level.
\end{abstract}

\begin{flushleft}
PACS: 11.10.Nx,
 11.25.Db

Keywords: noncommutative field theory, retarded functions

\end{flushleft}

\section{Introduction}

Noncommutative Quantum Field Theory (NCQFT) has recently received renewed
attention (see \cite{DouglasNekrasov} for a review). This interest is
triggered by its appearance in the context of string theory
 \cite{SeibergWitten}, and by the observation that Heisenberg's 
 uncertainty principle along with general relativity suggests the 
 introduction of noncommutative space-time \cite{DFR}. 
 Its mathematical foundations may also be found in
 Connes' formulation of noncommutative geometry,
 Moyal noncommutative field theory has been shown to be
 compatible with the latter one in the Euclidean case \cite{Gayral}. Moreover, it arises
 in the framework of deformation quantization \cite{Kontsevich}. \\

Coordinates are considered as noncommuting hermitian
operators $\hat{x}^\mu$, which satisfy the commutation relation
\begin{equation}
[\hat{x}^\mu,\hat{x}^\nu]=i\theta^{\mu\nu}\quad.
\end{equation}
We will assume the antisymmetric matrix $\theta^{\mu\nu}$ to be constant. The algebra of these noncommuting coordinate operators can be realized on functions on the ordinary Minkowski space by introducing the Moyal $\star$-product
\begin{equation}
(f\star g)(x)=e^{\frac{i}{2}\theta^{\mu\nu}\partial_\mu^\xi\partial_\nu^\eta}f(x+\xi)g(x+\eta)\Big|_{\xi=\eta=0}
\quad .
\end{equation}
To obtain a NCQFT from a commutative QFT, one replaces the ordinary product of field operators by the star product in the action. Due to the trace property of the star product, meaning that
\begin{equation}
\int dx ~(f_1\star ... \star f_n)(x)
\end{equation}
is invariant under cyclic permutations, the free theory is not affected and noncommutativity only appears in the interaction part. As an example, the interaction in noncommutative $\varphi_\star^3$-theory reads
\begin{equation}
S_{\text{int}}=\frac{g}{3!}\int dx~(\varphi\star\varphi\star\varphi)(x)\quad . \\
\end{equation}
 
A first suggestion for perturbation theory has been made in
 \cite{Filk}, where the Feynman rules for the ordinary QFT are only modified
  by the appearance of momentum-dependent phase factors at the vertices. These are of the form
 $e^{-ip\wedge q}$, with $p\wedge q=\frac{1}{2}p_\mu\theta^{\mu\nu}q_\nu$.
 In the case of only space-space-noncommutativity, i.e. $\theta^{0i}=0$, this approach leads to
 the
 UV/IR mixing problem, a renormalizable model has
 been suggested in \cite{Wulkenhaar}.  
 The general case of space-time noncommutativity, i.e. $\theta^{0i}\neq 0$,
 raises problems at an earlier stage due to the nonlocality of the star product, which involves 
time-derivatives to arbitrary high orders. It has been shown that the S-matrix is no longer unitary 
 as the cutting rules are violated \cite{GomisMehen}, the corresponding calculation
 involves only the tree level and the finite part of the one-loop-level. \\

To cure this problem, a different perturbative approach, TOPT, has been suggested for scalar theories in \cite{LiaoSiboldTOPT}. It mainly builds on the observation that for space-time noncommutativity time-ordering and star product of operators are not interchangeable, their order matters. Defining TOPT by carrying out time-ordering \emph{after} taking star products, a manifestly unitary theory is obtained.\\

However, further problems arise. The explicit violation of causality inside
the region of interaction was discussed in \cite{Grosse}, however, this
alone does
not spoil the consistency of the
formalism. 
In \cite{ORZ} it has been shown that Ward identities in NCQED are violated if TOPT is applied, which could be traced back to altered current conservation laws on the quantized level \cite{ReiDiplomarbeit}. Moreover, remaining Lorentz symmetry, i.e. Lorentz transformations which leave the noncommutativity parameter $\theta^{\mu\nu}$ invariant, is not respected by TOPT \cite{TOPTInv}.\\

To formulate a consistent perturbative approach to space-time
noncommutative theories is thus still a task to work on. One recent
suggestion building on the observation of violated remaining Lorentz
symmetry in TOPT has been made in \cite{HeslopSibold}, another one starts
from the Yang-Feldman equations \cite{BahnsDiss}. In this Letter, we want to
investigate the approach via retarded functions as introduced in the
commutative case in \cite{LSZ2} and further elaborated in \cite{Symanzik},
a pedagogical presentation may also be found in \cite{Rzewuski}. In this
formalism, retarded functions are used instead of time-ordered Green's functions, the
motivation is that the usage of the first ones allows an easier derivation of
unitarity and causality due to certain support properties of retarded
functions. We will extend
this approach in a natural way to noncommutative theories and investigate unitarity
as well as quantized equations of motion. The latter is motivated by its
similarity to current conservation laws: if classical equations of motion
are not altered on the quantum level also classical current conservation
laws will remain valid on the quantized level. We will find both unitarity
as quantized equations of motion to be disturbed in a specific way that
allows to modify the theory such that it is unitary and preserves the
classical equations of motion on the quantum level.

\section{The commutative case}

\subsection{Retarded functions and the generating functional}

We consider a field theory with a single hermitian field $\phi$ of mass $m$. The retarded products are then given by retarded multiple commutators of $\phi$:
\begin{equation}
R(x;x_1...x_n)= (-i)^n\sum_{\text{perm}}\vartheta(x^0-x^0_1)...\vartheta(x^0_{n-1}-x^0_n)\big[...[\phi(x),\phi(x_1)]...\phi(x_n)\big]
\label{ret_prod}
\end{equation}
where the summation is taken over all permutations of the $n$ coordinates $x_i$, $\vartheta$ denotes the step function. The support property $R(x;x_1,...,x_n)\neq 0$ only for $x^0\geq x^0_1,...,x^0_n$ is immediately clear from this definition.
The retarded functions are now defined as the vacuum expectation values of the retarded products:
\begin{equation}
r(x;x_1...x_n)=\langle 0|R(x;x_1...x_n)|0\rangle
\end{equation}

and with their help the S-matrix may be obtained by a reduction formula
as elaborated by H. Lehmann,
K. Symanzik and W. Zimmermann in \cite{LSZ2}, the amputation of external
legs works as usual through multiplication by the propagator. \\

To compute retarded functions we follow \cite{Rzewuski} and introduce the generating functional
\begin{align}
\mathcal{R}[j',j]=&\exp\Big\{2\int dx\sin\Big(\frac{1}{2}\frac{\delta}{\delta j(x)}\frac{\delta}{\delta\frac{\delta}{\delta j'(x)}}\Big)\int dy~ \mathcal{L}_{\text{int}}\Big(\frac{\delta}{\delta j'(y)}\Big)\Big\}\times \cr
&  \quad \exp\Big\{\int dzdw\Big(\frac{1}{4}j'(z)\Delta^{(1)}(z-w)j'(w)-j'(z)\Delta^{ret}(z-w)j(w)\Big)\Big\}
\label{ret_func}
\end{align}
where $\Delta^{ret}$ is a Green's function to the Klein-Gordon equation
\begin{equation}
\Delta^{ret}(x)=\lim_{\epsilon\rightarrow +0} \frac{-1}{(2\pi)^4}\int d^4k\frac{e^{-ikx}}{(k+i\epsilon)^2-m^2}
\end{equation}
with the support property $\Delta^{ret}(x)=0$ for $x^0<0$ and
$\Delta^{(1)}$ is given by
\begin{equation}
\Delta^{(1)}(x)=\frac{1}{(2\pi)^3}\int d^4k~ \delta(k^2+m^2)e^{-ikx}
\end{equation}
 being a solution to the homogeneous Klein-Gordon equation: $(\Box+m^2)\Delta^{(1)}(x)=0$.
\newline
Retarded functions are obtained by means of functional differentiation:
\begin{equation}
r(x;x_1...x_n)=\frac{\delta}{\delta j'(x)}\frac{\delta^n}{\delta j(x_1)...\delta j(x_n)}\mathcal{R}[j',j]\Big|_{j'=j=0}    \qquad . \label{ret_fcts}
\end{equation}

\subsection{Diagrammatic rules \label{dia}}

For later purpose, we want to write the outcome of equation (\ref{ret_fcts}) in the form of diagrams. Its lines will obviously carry $\Delta^{ret}$ or $\Delta^{(1)}$, and for
 $r(x;x_1,...,x_n)$ there will be endpoints $x, x_1,...,x_n$.\newline

To see which diagrams are allowed according to (\ref{ret_fcts}), we expand the first exponential in (\ref{ret_func}) in the example of $\mathcal{L}_{\text{int}}=g\phi^m$:
\begin{align}
\mathcal{R}[j',j]=1+&\sum_{n=1}^\infty \frac{1}{n!}\prod_{i=1}^n\int dy_i~2\sin\Big(\frac{1}{2}\frac{\delta}{\delta j(y_i)}\frac{\delta}{\delta\frac{\delta}{\delta j'(y_i)}}\Big)\int dz_i~g^m\frac{\delta^m}{\delta j'(z_i)^m} \times \cr
& \times\exp\Big\{\int dydz\Big(\frac{1}{4}j'(y)\Delta^{(1)}(y-z)j'(z)-j'(y)\Delta^{ret}(y-z)j(z)\Big)\Big\} \quad .
\end{align}
Recalling that
\begin{equation}
r(x;x_1...x_n)=\frac{\delta}{\delta j'(x)}\frac{\delta^n}{\delta j(x_1)...\delta j(x_n)}\mathcal{R}[j',j]\Big|_{j'=j=0}
\end{equation}
we see that $x$ is connected by $\Delta^{ret}(x-a)$ or $\Delta^{(1)}(z-a)=\Delta^{(1)}(a-z)$, the points $x_i$ are connected by $\Delta^{ret}(a_i-x_i)$; $a, a_i$ being some inner or outer points.

The $\frac{\delta}{\delta\frac{\delta}{\delta j'(y_i)}}$ in the $\sin$ can only act on $\frac{\delta^m}{\delta j'(z_i)^m}$, such that by expanding $\sin$ we can make the replacement
\begin{align}
&\int dy_i~2\sin\Big(\frac{1}{2}\frac{\delta}{\delta j(y_i)}\frac{\delta}{\delta\frac{\delta}{\delta j'(y_i)}}\Big)\int dz_i~g^m\frac{\delta^m}{\delta j'(z_i)^m} \cr
\equiv&~
2\sum_{j\leq [\frac{m-1}{2}]}g^m\int dz_i\frac{1}{(2j+1)!}\big(\frac{1}{2}\big)^{2j+1}\frac{\delta^{2j+1}}{\delta j(z_i)^{2j+1}}\frac{\delta^{m-2j-1}}{\delta j'(z_i)^{m-2j-1}}
\end{align}
such that at the vertex $z_i$ we have an odd power of $\frac{\delta}{\delta j(z_i)}$. As an incoming $\Delta^{ret}(a-z_i)$ at vertex $z_i$ can only be created by $\frac{\delta}{\delta j(z_i)}$ and vice versa, we find that the number of incoming $\Delta^{ret}$-functions at each vertex must be odd.\newline

One checks that there are no further restrictions to diagrams as the ones mentioned above, so we have found the diagrammatic rules for the retarded function $r(x;x_1...x_n)$:
\begin{enumerate}
\item $x,x_1,...,x_n$ are the endpoints of the diagram, inner points are called vertices.
\item $\Delta^{ret}(x-y)$ is symbolized by
\begin{picture}(50,10)(0,0)
\Vertex(5,5){1}
\Vertex(45,5){1}
\ArrowLine(5,5)(45,5)
\Text(5,-2)[]{$x$}
\Text(45,-2)[]{$y$}
 \end{picture},$~~$
  $\Delta^{(1)}(x-y)=\Delta^{(1)}(y-x)$ by
\begin{picture}(50,10)(0,0)
\Vertex(5,5){1}
\Vertex(45,5){1}
\Line(5,5)(45,5)
\Text(5,-2)[]{$x$}
\Text(45,-2)[]{$y$}
 \end{picture} ~~ .
\item $x$ is connected by one line, $\Delta^{ret}(x-a)$ or $\Delta^{(1)}(x-a)$. The points $x_i$ are also connected by one line each, $\Delta^{ret}(a_i-x_i)$.
\item The number of lines at each vertex is $m$ for $\phi^m$-theory, the contributing factor g, one integrates over the vertices.
\item The number of incoming functions $\Delta^{ret}(a_i-z_i)$ at each vertex $z_i$ is odd.
\end{enumerate}

\section{The noncommutative case\label{noncomm}}

We implement noncommutativity by defining retarded functions via the generating functional (\ref{ret_func}), where the interaction now involves the star product, e.g. in noncommutative $\phi^3_\star$-theory $S_{\text{int}}=\frac{g}{3!} \int dx (\phi\star \phi\star \phi)(x)$. This results in star multiplication at each vertex. In the Fourier representation of the retarded functions we thus encounter at every vertex a noncommutative phase factor $V(\pm p_1,...,\pm p_m)$ if $p_1,...,p_m$ are the momenta flowing
${\scriptsize \bigl\{
       \begin{aligned}&\text{in}\\
                 &\text{out}\end{aligned}\bigr\}}$
of the vertex. This phase factor is given by the m-point-function at first order, e.g.  in $\phi^3$-theory it reads
\begin{equation}
V(p_1,p_2,p_3)= \frac{1}{6}\sum_{\pi\epsilon S_3} e^{-i(p_{\pi(1)}, p_{\pi(2)}, p_{\pi(3)})} \qquad .
\end{equation}
Here we made use of the abbreviation
\begin{equation}
(p_1,...,p_n)=\sum_{i<j}p_i\wedge p_j
\end{equation}
where the $\wedge$-product is defined as $p\wedge q=\frac{i}{2}p_\mu\theta^{\mu\nu}q_\nu$.

In space-time noncommutative theories this way of introducing retarded
functions will not respect their support properties, i.e. in general we
will also outside the region $x^0\geq x^0_1,...,x^0_n$ have non-vanishing
$r(x;x_1,...,x_n)$. This is due to the fact that for $\theta^{0i}\neq 0$
the star product involves time-derivatives, such that one smears over  the
time coordinate. It is therefore clear that one can no longer obtain the
so-defined retarded functions from retarded products of the form
(\ref{ret_prod}), as they were originally introduced. However, we still
consider the theory worth to be further studied, and compute S-matrix
elements by using the reduction formula.\\

To obtain diagrammatic rules for the noncommutative case, the ones from the previous subsection only have to be supplemented by the rule
\begin{itemize}
\item[6.] At every vertex $x$ we perform star multiplication with respect to $x$.
\end{itemize}

\subsection{Unitarity}

To analyze unitarity, we follow closely the presentation in \cite{Rzewuski}.
 There the generalized unitarity condition
\begin{equation}
\mathcal{R}[0,j]=1
\label{uni_cond}
\end{equation}
is derived which implies unitarity for the S-matrix. The analysis of this condition in noncommutative theories will be the aim of this section.
We consider the case of $\phi^m$-theory and start with performing a Taylor expansion of the first exponential in (\ref{ret_func}):
\begin{align}
\mathcal{R}[0,j]=1+&\sum_{n=1}^\infty \frac{1}{n!}\prod_{i=1}^n\int dy_i~2\sin\Big(\frac{1}{2}\frac{\delta}{\delta j(y_i)}\frac{\delta}{\delta\frac{\delta}{\delta j'(y_i)}}\Big)\int dz_i~g^m\frac{\delta^m}{\delta j'(z_i)^m} \times \cr
& \times\exp\Big\{\int dydz~\Big(\frac{1}{4}j'(y)\Delta^{(1)}(y-z)j'(z)-j'(y)\Delta^{ret}(y-z)j(z)\Big)\Big\}\Bigg|_{j'=0}
\end{align}
Each factor in the n-th term ($n \geq1$) of the sum contains at least one functional derivative $\frac{\delta}{\delta (z_i)}$ such that we obtain $\prod_{i=1}^n \int dx_i j'(x_i)\Delta^{ret}(x_i-z_i)$ in front of the exponential, which does not vanish for $j'=0$ only if every factor is differentiated with some $\frac{\delta}{\delta j(z_j)}$. This means that at each vertex $z_i$ we have an ending $\Delta^{ret}(a-z_i)$, and the point $a$  must be again out of the $\{z_i\}_{i=1}^n$, which implies that we have a closed cycle of $\Delta^{ret}$-functions, i.e. an expression of the form
\begin{equation}
\begin{picture}(0,0)(0,0)
\LongArrow(5,-10)(5,-3)
\Line(5,-10)(259,-10)
\Line(259,-5)(259,-10)
\end{picture}
\Delta^{ret}(z_{1}-z_{2})\stackrel{z_2}{\star}\Delta^{ret}(z_{2}-z_{3})\stackrel{z_3}{\star}...\stackrel{z_k}{\star}\Delta^{ret}(z_{k}-z_{1})\stackrel{z_1}{\star} \quad . \label{star_cycle}
\end{equation}

The last statement can be seen as follows: choose $z_{i_1}$,
which appears in a function $\Delta^{ret}(a-z_{i_1}), a$ among
the $z_i$'s, say $a=z_{i_2}$. Either $z_{i_2}= z_{i_1}$ and we
have found a closed cycle, or $z_{i_2}\neq z_{i_1}$ in which case we
proceed  by finding $z_{i_3}$ such that $\Delta^{ret}(z_{i_3}-z_{i_2})$
appears. In the case $z_{i_3}= z_{i_1}$ or $z_{i_3}= z_{i_2}$ we are
finished, otherwise we go on in the same way. The limited number of
points $\{z_i\}_{i=1}^n$ implies that the procedure will stop and yield
a closed cycle of $\Delta^{ret}$-functions.\\
This means that the only terms which spoil the unitarity condition
(\ref{uni_cond}) contain a closed cycle of $\Delta^{ret}$-functions.
Let us first consider the case $\theta^{0i}=0$, where the star product
does not involve time derivatives. From the support property
\begin{equation}
\Delta^{ret}(x)\neq0 \quad \text{only for} \quad x^0>0
\end{equation}
we find as a condition that (\ref{star_cycle}) does not vanish
\begin{equation}
z_1^0>z_2^0>...>z_k^0>z_1^0
\end{equation}
which cannot be fulfilled, meaning that (\ref{star_cycle}) is zero. \\
However, in the general case of space-time noncommutativity, one can no
longer use this argumentation, as then taking star products contains a
smearing over the time coordinates. In fact, it was argued in
\cite{BahnsDiss}, that e.g.
$\Delta^{ret}(x)\star\Delta^{ret}(-x)\neq 0$. The diagrams involving
expressions (\ref{star_cycle}) thus are the ones which violate unitarity
if time does not commute with space. \\

\subsection{Composite operators: equations of motion and currents \label{comp}}

To derive equations of motion on the quantized level, i.e. on the level
of retarded functions, we define retarded functions
$r^\mathcal{O}(x;x_1...x_n)$ for a composite operator $\mathcal{O}$ at
place $x$ and single fields at $x_1,...,x_n$ in the following way. We
differentiate the generating functional by $\frac{\delta}{\delta j'(x)}$
once for every single field appearing in  $\mathcal{O}$ and by
$\frac{\delta^n}{\delta j(x_1)...\delta j(x_n)}$. For $\mathcal{O}$ in
the form $\mathcal{O}=D_1\phi \star D_2\phi\star...\star D_k\phi$ with
$D_i$ differential operators this means
\begin{align}
& r^{D_1\phi\star D_2\phi\star...\star D_k\phi}(x;x_1...x_n) \cr
& \equiv   D_1\frac{\delta}{\delta j'(x)}\star D_2\frac{\delta}{\delta j'(x)}\star...\star D_k\frac{\delta}{\delta j'(x)}\frac{\delta^n}{\delta j(x_1)...\delta j(x_n)}\mathcal{R}[j',j]\Big|_{j'=j=0}
\end{align}
e.g.
\begin{equation}
r^{\phi \star(\Box+m^2)\phi}(x;x_1...x_n) \equiv \frac{\delta}{\delta j'(x)}\star(\Box_x+m^2)\frac{\delta}{\delta j'(x)}\frac{\delta^n}{\delta j(x_1)...\delta j(x_n)}\mathcal{R}[j',j]\Big|_{j'=j=0} \quad .
\end{equation}

Diagrammatic rules for $r^\mathcal{O}(x;x_1...x_n)$ with
$\mathcal{O}=D_1\phi \star D_2\phi\star...\star D_k\phi$  can be
easily read off, the only change to the previous rules lies in how
the point $x$ is treated, we therefore replace rule 3 by
\begin{itemize}
\item[3'.] $x$ is connected by $k$ lines; the $i^{\text{th}}$ line
carries $D_i\Delta^{ret}(x-a_i)$ or $D_i\Delta^{(1)}(x-a_i)$. The points
$x_i$ are connected by one line each, $\Delta^{ret}(b_i-x_i)$. Star
multiplication with respect to $x$ is performed.
\end{itemize}

As an example of equations of motion and current conservation laws we
now want to prove the bilinear equation of motion in
$\phi_\star^3$-theory, which classically reads
\begin{equation}
\phi\star(\Box+m^2)\phi=g\phi\star\phi\star\phi \quad ,
\end{equation}
on the level of retarded functions, i.e. show that
\begin{equation}
r^{\phi\star(\Box+m^2)\phi}(x;x_1...x_n)=r^{g\phi\star\phi\star\phi}(x;x_1...x_n)+ \text{c.t.}   \label{eqmo}
\end{equation}
with c.t. meaning contact terms.
We will evaluate both sides of the above equation diagrammatically:
\begin{align}
r^{\phi \star(\Box+m^2)\phi}(x;x_1...x_n)=&
\begin{picture}(100,50)(0,40)
\Vertex(50,85){1}
\Vertex(30,0){1}
\Vertex(70,0){1}
\GOval(50,42.5)(17,30)(0){0.75}
\DashArrowLine(50,85)(30,55){2}
\DashArrowLine(50,85)(70,55){2}
\ArrowLine(30,30)(30,0)
\ArrowLine(70,30)(70,0)
\Text(50,95)[]{$x$}
\Text(30,-10)[]{$x_1$}
\Text(50,-10)[]{...}
\Text(70,-10)[]{$x_n$}
\Text(67,75)[l]{$(\Box+m^2)$}
\end{picture}
 \cr \cr \cr \cr
=&\int dy
\begin{picture}(100,70)(10,40)
\Vertex(50,85){1}
\Vertex(30,0){1}
\Vertex(70,0){1}
\Vertex(60,70){1}
\GOval(50,42.5)(17,30)(0){0.75}
\DashArrowLine(50,85)(30,55){2}
\ArrowLine(50,85)(60,70)
\ArrowLine(30,30)(30,0)
\ArrowLine(70,30)(70,0)
\DashLine(60,70)(50,60){2}
\DashLine(60,70)(70,55){2}
\Text(50,95)[]{$x$}
\Text(70,70)[]{$y$}
\Text(30,-10)[]{$x_1$}
\Text(50,-10)[]{...}
\Text(70,-10)[]{$x_n$}
\Text(60,85)[l]{$(\Box+m^2)$}
\end{picture} +
\quad \sum_{k=1}^n
\begin{picture}(100,70)(0,40)
\Vertex(65,85){1}
\Vertex(30,0){1}
\Vertex(70,0){1}
\Vertex(110,0){1}
\GOval(50,42.5)(17,30)(0){0.75}
\DashArrowLine(65,85)(30,55){2}
\ArrowLine(65,85)(110,0)
\ArrowLine(30,30)(30,0)
\ArrowLine(70,30)(70,0)
\Text(65,95)[]{$x$}
\Text(30,-10)[]{$x_1$}
\Text(50,-10)[]{...}
\Text(50,0)[]{$\check{x_k}$}
\Text(70,-10)[]{$x_n$}
\Text(110,-10)[]{$x_k$}
\Text(93,50)[l]{$(\Box+m^2)$}
\end{picture} \nonumber
\end{align}
\\ \\ \\

where the dashed arrow line
\begin{picture}(50,10)(0,0)
\Vertex(5,5){1}
\Vertex(45,5){1}
\DashArrowLine(5,5)(45,5){2}
 \end{picture}
can be
\begin{picture}(50,10)(0,0)
\Vertex(5,5){1}
\Vertex(45,5){1}
\ArrowLine(5,5)(45,5)
 \end{picture}or
\begin{picture}(50,10)(0,0)
\Vertex(5,5){1}
\Vertex(45,5){1}
\Line(5,5)(45,5)
\end{picture}
and the dashed line
\begin{picture}(50,10)(0,0)
\Vertex(5,5){1}
\Vertex(45,5){1}
\DashLine(5,5)(45,5){2}
 \end{picture}
stands for
\begin{picture}(50,10)(0,0)
\Vertex(5,5){1}
\Vertex(45,5){1}
\ArrowLine(5,5)(45,5)
 \end{picture},
\begin{picture}(50,10)(0,0)
\Vertex(5,5){1}
\Vertex(45,5){1}
\ArrowLine(45,5)(5,5)
 \end{picture}
 or
\begin{picture}(50,10)(0,0)
\Vertex(5,5){1}
\Vertex(45,5){1}
\Line(5,5)(45,5)
\end{picture}.
We have used $(\Box+m^2)\Delta^{(1)}(x)=0$ to skip diagrams
that have a line $\Delta^{(1)}$ between $x$ and $y$ resp. $x$
and $x_k$. \\Applying   $(\Box+m^2)\Delta^{ret}(x)=\delta(x)$
we recognize the last diagram as contact terms, such that
\begin{align}
r^{\phi\star (\Box+m^2)\phi}(x;x_1...x_n)=&
\begin{picture}(100,70)(0,40)
\Vertex(50,85){1}
\Vertex(30,0){1}
\Vertex(70,0){1}
\GOval(50,42.5)(17,30)(0){0.75}
\DashArrowLine(50,85)(30,55){2}
\DashLine(50,85)(70,55){2}
\DashLine(50,85)(50,60){2}
\ArrowLine(30,30)(30,0)
\ArrowLine(70,30)(70,0)
\Text(50,95)[]{$x$}
\Text(30,-10)[]{$x_1$}
\Text(50,-10)[]{...}
\Text(70,-10)[]{$x_n$}
\end{picture}
\quad + \quad \text{c.t.} \label{eqmol}
\end{align}
\\
\\ \\ \\
The right hand side of equation (\ref{eqmo}) yields in terms of diagrams
\begin{align}
r^{g\phi\star\phi\star\phi}(x;x_1...x_n)=&
\begin{picture}(100,70)(0,40)
\Vertex(50,85){1}
\Vertex(30,0){1}
\Vertex(70,0){1}
\GOval(50,42.5)(17,30)(0){0.75}
\DashArrowLine(50,85)(30,55){2}
\DashArrowLine(50,85)(70,55){2}
\DashArrowLine(50,85)(50,60){2}
\ArrowLine(30,30)(30,0)
\ArrowLine(70,30)(70,0)
\Text(50,95)[]{$x$}
\Text(30,-10)[]{$x_1$}
\Text(50,-10)[]{...}
\Text(70,-10)[]{$x_n$}
\end{picture}
\end{align}
\\ \\
To investigate under which conditions both sides are equal up to contact
terms, we need to analyze under which conditions diagrams belonging to
(\ref{eqmol}) with a dashed line being a $\Delta^{ret}$-function that
points to $x$ are zero. At first, we prove the following
\begin{lemma}
 A diagram having at each vertex at least one incoming
 $\Delta^{ret}$-function attached and the endpoints connected by
 outgoing $\Delta^{ret}$-functions contains a closed cycle of
 $\Delta^{ret}$-functions.
\end{lemma}
\emph{Proof:} Let $\{z_i\}_{i=1}^n$ be the set of vertices, at
each $z_i$ we have a function $\Delta^{ret}(a_i-z_i)$, and $a_i$ must,
as the outer points are connected by outgoing $\Delta^{ret}$-functions,
be itself out of $\{z_i\}_{i=1}^n$. We can now use the same argumentation
as in the discussion of unitarity to obtain a closed cycle of
$\Delta^{ret}$-functions.\\

If we consider the point $x$ not as an endpoint but a vertex of the
diagram, we find that diagrams belonging to (\ref{eqmol}) with a dashed
line being a $\Delta^{ret}$-function that points to $x$ contain a closed
cycle of $\Delta^{ret}$-functions.
From our discussion of closed cycles of $\Delta^{ret}$-functions in the
previous section we know that these vanish for $\theta^{0i}=0$ but not
necessarily otherwise. We have thus found that the classical bilinear
equation of motion holds on the quantum level in the case of only spatial
noncommutativity. However it will be disturbed by diagrams containing
closed cycles of $\Delta^{ret}$-functions if time does not commute with
space. This results generalizes to quantum current conservation laws,
which are derived in a  similar manner.

\subsection{A modified theory}

Let us first summarize our results so far. For space-time noncommutativity
unitarity has turned out to be violated and the classical equations of
motion and currents do not hold on the quantized level. In both cases
these unpleasant outcomes are exactly due to diagrams which contain a
closed cycle of $\Delta^{ret}$-functions. Their vanishing for
$\theta^{0i}=0$ is the reason that in this case the approach via
retarded functions yields a unitary theory and respects the classical
equations. \\
The motivation to modify the theory is to obtain a theory which is unitary 
and preserves the classical equations of motion, therefore current
conservation laws, on the tree-level and the finite part of the
one-loop-level.\\
With the above results, it is obvious that we encounter these properties if we
alter the theory by the requirement that we do not allow diagrams which
exhibit a closed cycle of $\Delta^{ret}$-functions. This modified theory
can probably not be derived from a functional like (\ref{ret_func}),
instead it is  defined by the diagrammatic rules of subsection \ref{dia}
together with the rules of section \ref{noncomm} and subsection \ref{comp}
if we impose the additional requirement
\begin{itemize}
\item[7.] A diagram must not contain a closed cycle of
$\Delta^{ret}$-functions.
\end{itemize}
As diagrams which are excluded by the above rule vanish for $\theta^{0i}=0$
the equivalence of the
modified theory with the ordinary one derived from (\ref{ret_func})
in the case of only spatial noncommutativity is evident.

Let us briefly comment on Lorentz covariance: each diagram only involves
expressions which are covariant under Lorentz transformations (if we also
transform $\theta^{\mu\nu}$), thus are
Lorentz covariant. This property is therefore not disturbed by excluding a
certain type of diagrams, meaning that the modified theory is still Lorentz
covariant. We will thus expect it to respect remaining Lorentz symmetry.

\section{Conclusions}

We have extended retarded functions to noncommutative quantum field theories
and analyzed the resulting perturbation theory. In space-time noncommutative
theories we have found that unitarity is violated and the classical equations
of motion and currents are not respected on the quantum level.
Both unpleasant results can be ascribed to the same type of diagrams,
which vanish in the case of only spatial noncommutativity. Modifying
the theory by explicitly forbidding them yields a theory which has the
desired properties of being unitary and respecting classical equations
of motion and currents on the quantum level. This theory is defined by
a set of diagrammatic rules, for vanishing $\theta^{0i}$ it coincides with
the unmodified approach.

\section*{Acknowledgements}

I wish to thank K. Sibold and C. Dehne for the incitations to study retarded
functions in the context of space-time noncommutative theories and for
fruitful
discussions on this work.

\end{document}